# Insider trading in the run-up to merger announcements.

# Before and after the UK's Financial Services Act 2012.

## Rebecca Pham[1,a] and Marcel Ausloos[1,b,2,3,*]

[1] School of Business, College of Social Sciences, Arts, and Humanities, University of Leicester, Leicester, LE2 1RQ, United Kingdom

email:

(a) rebeccapham@hotmail.co.uk; rp409@student.le.ac.uk;

(b) ma683@le.ac.uk

[2] Department of Statistics and Econometrics, Bucharest University of Economic Studies, Calea Dorobantilor 15-17, Bucharest, 010552 Sector 1, Romania

email: marcel.ausloos@ase.ro

[3] Group of Researchers for Applications of Physics in Economy and Sociology (GRAPES), Rue de la belle jardinière, 483, Sart Tilman, B-4031 Angleur, Liège, Belgium

email: marcel.ausloos@ulg.ac.be

- corresponding author
- **ORCID: 0000-0001-9973-0019**

We confirm that this article contains a Data Availability Statement (see Section 4).

We confirm that we include how available data was and can be obtained.

## Abstract

After the 2007/2008 financial crisis, the UK government decided that a change in regulation was required to amend the poor control of financial markets. The Financial Services Act 2012 was developed as a result in order to give more control and authority to the regulators of financial markets. Thus, the Financial Conduct Authority (FCA) succeeded the Financial Services Authority (FSA). An area requiring an improvement in regulation was insider trading. Our study examines the effectiveness of the FCA in its duty of regulating insider trading through utilising the event study methodology to assess abnormal returns in the run-up to the first announcement of mergers. Samples of abnormal returns are examined on periods, under regulation either by the FSA or by the FCA. Practically, stock price data on the London Stock Exchange from 2008-2012 and 2015-2019 is investigated. The results from this study determine that abnormal returns are reduced after the implementation of the Financial Services Act 2012; prices are also found to be noisier in the period before the 2012 Act. Insignificant abnormal returns are found in the run-up to the first announcement of mergers in the 2015-2019 period. This concludes that the FCA is efficient in regulating insider trading.

# 1. Introduction

Trading has been a vital mechanism in economies for centuries. In the present days, one of the most popular methods of trading for monetary gain occurs through the transactions in financial markets (most particularly through stock exchange markets). Trading is legal and most importantly only fair when all parties have access to the same information. The phenomenon of insider trading occurs in the presence of asymmetrical information, when one party has access and acts upon private information, trading on securities with the unfair advantage to make abnormal profits. This generates problems like agency costs between managers and shareholders. Consequently, insider trading legislation has been enforced from as early as 1933 in the US Securities Act (Thompson, 2013).

Although insider trading is now illegal in most countries, a major debate stands concerning whether markets require the presence of insider trading to operate efficiently. Critics of legislation state that the presence of insider trading in markets generates more efficient prices (Manne, 1966). However, those in favour of illegalising insider trading express their concerns about insider trading causing a misallocation of resources within markets (Kim et al., 2019).

Scholars in this area of research have found abnormal returns to be significant in the run-up to events, specifically in US stock markets. To expand on this literature, we focus on securities on the London Stock Exchange (*https://www.londonstockexchange.com/*). Insider trading has been illegal in the UK since 1980 with regulation in force since 2001 as a result of the Financial Services and Markets Act 2000 (Thompson, 2013). After the 2007/2008 financial crisis, criticisms were received by the regulatory authority, the Financial Services Authority (FSA). Consequently, the UK government established the Financial Services Act 2012 to change the regulatory bodies of financial markets from the FSA to the Financial Conduct Authority (FCA) (Financial Services Act, 2012*)*. In comparison to the FSA where both customer interests and the risks of institutions were managed, the FCA was created to focus solely on customer interests (Anker, 2013). However, criticisms are still received by the FCA due to their lack of insider trading convictions (Chapman, 2018). Thus, it is of interest to examine the extent of insider trading in the UK before and after the Financial Services Act 2012 to assess the performance of the FCA. To do this, we investigate abnormal returns prior to merger and acquisition (M&A) announcements using an "event study methodology"

An M&A can be executed in either hostile or agreed conditions but in most cases is carried out with aims to expand. The planning required to reach a successful agreed merger involves months of collaboration from both companies. This involves numerous people who have information about these planned events months prior, so insider trading is a probable occurrence, as for example discussed by Agarwal and Singh (2006) or Kiriacou et al. (2010). Beside, it has been observed that over 10% of common stocks are held by insiders (Lorie and Niederhoffer, 1968).

The paper goes as follows: firstly, the literature review (Section 2) involves a description of the background of insider trading, next the discussion of insider trading regulation, and thereafter the consequence of insider trading for M&A events on asset pricing theory. This suggests research questions (Section 3) considered as a gap in the literature, but specifically with the UK case in mind (Section 4). The UK selected data is outlined and classically analysed, - connecting and discussing theoretical and empirical aspects pointed out in the literature review (Section 5). Finally, the conclusion (Section 6) provides a summary of this study, its limitations, and recommendations for future research.

# 2. Literature review

This section organization is of major importance in order to outline aspects of insider trading. Firstly, we discuss the background of the topic, including motivations as to why insiders trade. Next, we examine the legislation and regulation surrounding insider trading and the current debate behind this. Finally, we introduce the problem of insider trading on asset theories. Research questions are established in the following section from this brief review of the existing literature; we restrict the citations to what are considered the most pertinent papers for our purpose. Notice a more general review in Doffou (2003).

## 2.1 Informativeness

Adverse effects of insider trading over the informativeness of prices come from Rozeff and Zaman (1988) who find that outsiders mimicking insiders to gain abnormal returns create an incorrect pricing of securities, whence inefficiencies in the market. Fishman and Hagerty (1992) explain that due to insider trading, outsiders are prevented from acquiring information and thus cannot trade efficiently.

In contrast, Lakonishok and Lee (2001) argue that insider trading in investments with long term horizons are informative, - due to an under-reaction of the market. Kim et al., (2019) understand that this could be due to the differences in investment horizons rather than informativeness of stock prices. Kim et al., (2019) determined that allowing insiders to take advantage of inside information leads to noisier stock prices. This shows that insider trading laws must exist in the prevention of the market failing as a communication system (Bhattacharya and Spiegel, 1991).

## 2.2 Profitability

Studies exploring the extent of profitability from insider trading come from authors such as Jaffe (1974) who found evidence to show that insiders make abnormal returns on trades in their own companies. Thereafter, Finnerty (1976) and Lin and Howe (1990) found further evidence observing that insiders recognise the profitable and unprofitable positions within their companies. Kim et al. (2019) investigated insider trading in 44 countries and discovered that insider trading is unprofitable in Norway and Hong Kong, but one can gain significant abnormal profits in Canada, The Netherlands, Italy and South Africa.

Notice that Manne (1966) argues that profits attained from insider trading are necessary in order to contribute to the economic process.

**2.3 Managerial incentives**

Another important area of discussion in the literature of insider trading concerns the managerial incentives that motivate an insiders' actions and the effects this has on shareholders and the wider market. First, insider trading has been said to improve the welfare of the manager and the firm's shareholders (Dye, 1984). It is admitted that a rational shareholder will act immediately to new information confirmed in a merger announcement (Rosen, 2006); hence, once an insider receives private information on a merger, it is beneficial to act on this and trade illegally. An astute shareholder would follow up the move as quickly as possible.

Next, why managers may decide to act on inside information includes diversifying their portfolio (Ma et al., 2000), avoiding earnings disappointment (Richardson et al., 2010), increasing their stake in the firm (Hu and Noe, 1997), and maximising the equity of their firm (Bradley and Seyhun, 1997). These authors also determined that trading just before the release of information is a way for managers to undo the effects of insider holdings.

For completeness, notice that such managerial incentives are commonly explained by two theories: either agency theory or market theory. In brief, the agency theory focuses on the conflict of interest between managers and shareholders in which moral hazard and agency costs are important aspects. The market theory differs from agency theory as this relates to the wider market where it is found that permitting insider trading can induce managers to increase their stake in the firm's equity (Hu and Noe, 1997).

Convincingly, the agency theory explains the effects of insider trading on stockholders. Bradley and Seyhun (1997) said that the incentives of corporate managers and stockholders are more aligned when managers hold stock in firms they manage. This is because both parties have the common objective to make a profit from the relating stock/firm. Thus, allowing managers to trade on inside information can improve the welfare of shareholders (Bradley and Seyhun, 1997). Indeed, if the goals of the managers and shareholders are misaligned, this creates agency costs. Fox (1999) found that further agency costs that arise from inaccurate prices do lower liquidity and reduce the shareholder's motivations to monitor prices, whence increases insider's ability and incentive to expropriate outside investors. Of course, agency costs are reduced if insider trading is reduced and if the conflict between parties is reduced (Carlton and Fischel, 1983).

Differing from agency theory, the market theory focuses on stock price efficiency in the market. This is of interest here because insider trading can be viewed as a signalling activity based on private information, since prices will adjust to reflect the private news. Thus, investment and financing decisions of firms may be influenced by the security holdings of their managers. However, although signalling allows outsiders to also benefit from this information by analysing stock markets, this is worthless if outsiders do not know how to interpret this information, so prices are not always accurate in the market, and inefficiencies are created. In fact, Klock (1994) found that allowing managers to trade on inside information might give them incentives to take on too much risk or to undertake value reducing projects, reducing the value of the firms and creating inefficient firms. Recently, some scholars argue that insider trading is necessary not only to promote market efficiency but also to provide a good compensation scheme for managers (Bainbridge, 2019).

**2.4 UK Insider dealing regulation**

UK legislation first criminalised insider trading in 1980 through the UK Companies Act (1980) in sections 69-73 (Wijesinghe, 2018). After this, the Criminal Justice Act (1993) was introduced to define insiders and the illegality of insider trading. In 2000, the Financial Services and Markets Act was first created to identify the regulators of this legislation and their duties (Wijesinghe, 2018). The Criminal Justice Act (1993) on insider dealing in the UK states that it is illegal when an individual acts on private information or encourages another person to do so. Defying this act can lead to imprisonment of up to 7 years and/or a fine. If the defendant can prove the profit earned was unexpected, there is potential to be proven innocent and let off any charges. Under UK law, insider information is related to information that concerns a particular security, is specific/precise, and is not public news.

The role of regulating UK banks is currently assigned to two main regulators: the Prudential Regulation Authority (PRA) and the Financial Conduct Authority (FCA), which are both overseen by the Bank of England (Hall, 2013). The PRA and the FCA are the regulating bodies for banks, in which the PRA is the prudential regulator and the FCA is the conduct regulator. These two bodies succeeded the Financial Services Authority (FSA) in 2013 as a requirement of the Financial Services Act (2012). The Financial Services Act 2012 called for the takeover of the FSA for a system with better communications after the FSA's poor control during the 2007/2008 financial crisis (BBC News, 2013). The FCA is now responsible for "conduct of business, effective financial markets, consumer protection, and for promoting effective competition" (Practical Law, 2018). In this, a duty of the FCA is to investigate cases involving insider trading and provide evidence to incriminate the guilty. Despite convictions, figures have shown that almost a fifth of takeovers is preceded by suspicious share price movements (Chapman, 2018). Alongside this, the FCA (Financial Conduct Authority, 2016) discovered that insider trading occurs in about 10% of takeover announcements. Although this has been at its lowest level for 12 years, the number of cases which the FCA must tackle is increasing. To deter offenses, fines have been increased by over 220% (Binham, 2019). But despite this, there are still criticisms that the responses to insider trading by the FCA are too slow: a mere 12 convictions secured in 5 years (Chapman, 2018). This lack of solved cases demonstrates inefficiencies with the processes the regulatory body follows.

## 2.5 The debate for changes in regulations

The enforcement of insider dealing laws only became popular in the 1990s (Bhattacharya and Daouk, 2002) but the need to tighten regulations has been ongoing and is motivated by the notion that insider trading is unfair, unethical, and most importantly undermines public confidence in capital markets (Carlton and Fischel, 1983). Block and Smith (2015) explained that a loss in market confidence deters potential investors and cause equity markets to be less liquid. The deterrence of potential investors would cause a misallocation of investment in the market and hence result in less efficient markets. Due to this, shareholders may demand a higher return to compensate for the difficulty in analysing firms (Manne, 1966).

An investigation of insider trading regulation and stock price informativeness was conducted by Fenandes and Ferreira (2009) who concluded that the first enforcement of insider trading regulation improves stock price informativeness. Additional investigation of insider trading regulation and informativeness of stock prices was carried out by Kim et al. (2019). These authors determined that insider trades are significantly more informative in countries with active enforcement of insider trading; this informativeness leads to greater efficiency and allocation in financial markets. In regards to insider trading acting as a compensation scheme for managers, the discussion of insider trading regulation by Carlton and Fishcel (1983) raised another requirement to tighten regulation due to insider trading being an inefficient compensation scheme as this would only benefit risk-averse managers.

Block and Smith (2015) stated that the benefits that come from insider trading include improved price stability, increased transparency, earlier fraud detection, and is also an overall effective compensation scheme for managers. With this, Manne (1966) explains that there is an increase in price accuracy in the market when insider trading is present as the price of the affected security would be moving towards the correct price; where the correct price is the price that the security would be when all relating information is disclosed. Manove (1989) also discussed that the presence of insider trading brings private information to the market, allowing for the efficient allocation of resources. This is supported by Bushman et al. (2005) who studied data from 100 countries to conclude that insider trading restrictions lead to increased ''information acquisition effects'': individuals obtain information with greater ease thereby leading to the efficient allocation of information and resources. To further support this, an assessment of the Hong Kong stock market portfolio after revoking the new blackout regulation was conducted by Chen et al. (2018). These authors determined that insider trading is necessary to reduce the amount of asymmetrical information between insiders and outsiders. Alongside this, the authors discovered that enforcing tighter insider trading laws could bring about unintended consequences to shareholders and hurt the shareholder value.

## 2.6 Further theoretical framework

### 2.6.1 The random walk theory

A prominent theory in financial markets is known as the random walk (RW) theory (Darrat and Zhong, 2000). This states that the movement of asset prices is completely random and consequently cannot be predicted (Kendall and Hill, 1953). The presence of insider trading creates inconsistencies in the random walk theory as stock prices are no longer random and can be predicted by insiders. To disprove the RW theory, Wuttidma (2015) established that a positive relationship between aggregate insider trading and stock market volatility is present. This is due to the increased flow of information to the stock market resulting from insider trading, thereby increasing stock market volatility. In this study, it was also revealed that future stock market returns can be predicted from medium insider trades using information about expected cash flows and discount rate news which should be deemed impossible under the RW theory.

Studies focusing on merger and acquisition announcements and stock price movements have found significant evidence of market reactions before the first announcement of the merger. An investigation conducted by Keown and Pinkerton (1981) discovered that although the market appears to adjust immediately to the first public announcement, half of this adjustment actually occurs before the announcement of the merger. Their findings show leakage of information up to 12 days prior to the first announcement. They explained the difficulty in keeping these announcements secret due to the sheer number of people involved in these procedures. This concludes that with the presence of looming news especially regarding mergers, stock price movements will be affected accordingly.

### 2.6.2 The efficient market hypothesis

Consistent with the RW theory is the efficient market hypothesis (EMH) (Fama, 1970; Jovanovic, 2018): it states that asset prices should fully reflect all available information, hence providing equal chances for all investors. Fama proposed three types of efficiency; these include weak form efficiency, semi-strong form efficiency, and finally strong form efficiency. The weak form efficiency integrates the least information, in which asset prices only reflect past prices. As the market forms become stronger, asset prices reflect more information. Hence, the semi-strong form incorporates both past prices and public information, and finally asset prices in the strong form efficiency reflect past prices, public information, and private information.

Despite the popularity of this theory, an evident dilemma arises in the presence of insider trading leading numerous scholars to find inconsistencies with the EMH. For example, Laffont and Maskin (1990) stated that when a private information is used by a trader big enough to affect prices, the presumed information efficiency of the price necessarily breaks down. Concerning the strong form efficiency, Baesel and Stein (1979) found inconsistencies because their data revealed that insiders and bank directors would necessarily earn positive returns. Despite this, they explained that these returns were in fact due to an inefficiency of the market spreading information rather than being a result of the inefficiency of the financial markets, whence still giving some ''robustness'' to the EMH. Keown and Pinkerton (1981) present evidence to support the semi-strong form efficiency in presence of insider trading. They discovered that the market reaction to new information is complete by the day of the announcement, - thus showing no signs of inconsistencies with this form in the EMH. However, this is not supported by Rozeff and Zaman (1988) after introducing outsiders into the study. With the presence of outsiders, the semi-strong form market efficiency is disrupted as the opportunity to mimic the actions of insiders and earn abnormal profits is presented to them. Hence, like the strong form, the semi-strong form efficiency can be disproved with the presence of insider trading, - but the conclusion might depend on the ''time horizon''.

Nevertheless, the evidence presented by all these authors shows that the EMH is conceptually an unrealistic model as it assumes that all investors are rational and also does not account for asymmetrical information. An improved model, suggested by Jensen (1978), considers that prices reflect information in which marginal benefits do not exceed the marginal costs. The efficiency level, therefore, further depends on the information set incorporated in the stock prices.

Of course, it is well known that the primary problem of the EMH is the assumption that all participants have access to the same information; obviously, this assumption does not hold in presence of insider trading. Hence, another development of the asset pricing theory builds on the determinants of market efficiency. In the EMH, the information available is the prime determinant but Rees (1995) establishes 5 other features of market efficiency alongside information availability. These include the number of participants in the market, low transaction costs, location independence, homogeneity, and competitive analysis. This is of great interest; however, we will not extend our analysis to a study of the information content and spreading for M&A events at such times.

## 3. Research Questions

Our investigation, on UK cases, is in line with the research conducted by Keown and Pinkerton (1981) on merger announcements and insider trading activity on the New York and American stock exchanges between 1975 and 1978. Using event study methodology, these authors found that half of the market reaction occurs before the first announcement; the remaining reaction occurs the following day if the information is released after the market closes. Another similar study conducted by Jain and Sunderman (2014) examined stock price movements around merger announcements. This investigation focused on emerging markets, specifically India from 1996 to 2010, where data was gathered from the Bombay stock exchange. The results supported those found in the former paper, whence showed that information held by insiders will make its way into the security price. Further analysis of mergers between different sectors found insider trading insignificant. This was explained by the lack of understanding managers have of other sectors. Alongside this, the analysis found that hostile mergers between different sectors led to insignificant insider dealing. On the Indian market, we already mentioned work by Agarwal and Singh (2006).

The question was also examined on the Australian market, between 2000 and 2009, by Aspris et al. (2014) who found that no significant pre-bid run-up ahead of takeover announcements.

As mentioned, another important element in the topic of insider dealing surrounds the legislation and regulations. Consequently, the present study examines the extent for which insider trading is present in the run-up to merger announcements in two periods, involving 2008 to 2012 and 2015 to 2019, to examine the effectiveness of the FCA following the 2012 Financial Services Act. Hence, the research questions are

Research Question 1: To what extent are abnormal returns significant in the run of merger announcements in the period 2008 – 2012, under regulation by the FSA?

Null hypothesis ($H_0$): abnormal returns are not significant.

Research Question 2: To what extent are abnormal returns significant in the run up of merger announcements in the period 2015-2019, under regulation by the FCA?

Null hypothesis ($H_0$): abnormal returns are not significant.

The overall aim is to test the significance of insider trading in the UK between the two periods in the run-up to merger and acquisition announcements, in particular as a result of the 2012 Financial Services Act. From this, we discuss whether there has been a beneficial change and whether one should reconsider the insider trading regulation in the UK.

## 4. Methodology

The research questions established in the previous section are investigated using quantitative secondary data. The data is based on information on M&A involving UK companies, gathered from the UK Office for National Statistics. The exact date of the first announcement of each merger is taken from formal letters of confirmation from The Financial Times (*https://www.ft.com/mergers-acquisitions*), The London Stock Exchange (*https://www.londonstockexchange.com/)*, or the corresponding company's website. The corresponding stock price data ranges for 120 days around the announcement date and are obtained from Bloomberg terminals (*https://www.bloomberg.com/professional/solution/bloomberg-terminal/* a fifth of takeovers is preceded and Yahoo finance where the adjusted closing price are used to account for dividends.

For being coherent, and due to data availability, we reduced the stock price data from 58 M&A announcements in the London Stock Exchange along with the corresponding stock price from the London Stock exchange of the same period to act as a control variable; 18 of these merger announcements come from the 2008-2012 period. The remaining 40 merger announcements occur during 2015-2019. Observe that the sample of merger announcements used in the 2008-2012 period consists of just under half of those used for the 2015-2019 period. Notice also that we have disregarded companies involved in mergers that had taken place after the involvement of an obvious disposal, - because of possible anticipation effects on stock price. The list of mergers and acquisitions used as well as their announcement date are included in the Supplementary Files as Table 1 and Table 2.

The date of the first announcement corresponds to the event date; it is regarded as day 0 in each case. Overall, the stock price data covers 120 days around the event date; this includes prices from 90 days prior to 30 days post the event. The total amount of stock index data is 54 and the total observations are 6960.

The model is given by:

$$R_{jt} = \alpha_{jt} + \beta_{jt} R_{mt} + E_{jt} \quad (1)$$

where

$R_{jt}$ = return for the stock *j* on day *t*; see formula (2) below

$R_{mt}$ = return for the whole market index on day *t*; see formula (3) below

$\alpha_{jt}$ = so called alpha, "constant of the regression analysis" of the stock *j* on day *t*

$\beta_{jt}$ = so called beta, regression analysis coefficient of the stock *j* on day *t*

$E_{jt}$ = estimated abnormal return for stock *j* on day *t*.

In this model, the return on stock is defined as:

$$R_{jt} = \frac{(P_{jt} - P_{jt-1})}{P_{jt-1}} \quad (2)$$

where

$P_{jt}$ = price of the stock *j* on day *t*.

Similarly, the corresponding return on the market is calculated with:

$$R_{mt} = \frac{(P_{mt} - P_{mt-1})}{P_{mt-1}} \quad (3)$$

where

$P_{mt}$ = value of the market index on day *t*.

There is no need to account for dividends in this calculation as adjusted closing prices are used.

In event study methodology, different time windows are involved: the "estimation window" and the "event window". The estimation window is defined as the period of time before the event date where prices are not affected by the event in question, so abnormal return can be calculated using a "normal" price. The estimation window is used for the calculation of the alpha and beta parameter values in the market model; this window must be determined for each security. The event window involves the period in which abnormal returns are likely to occur and thus are tested for significance.

The timeline of an event so studied is shown in Fig. 1.

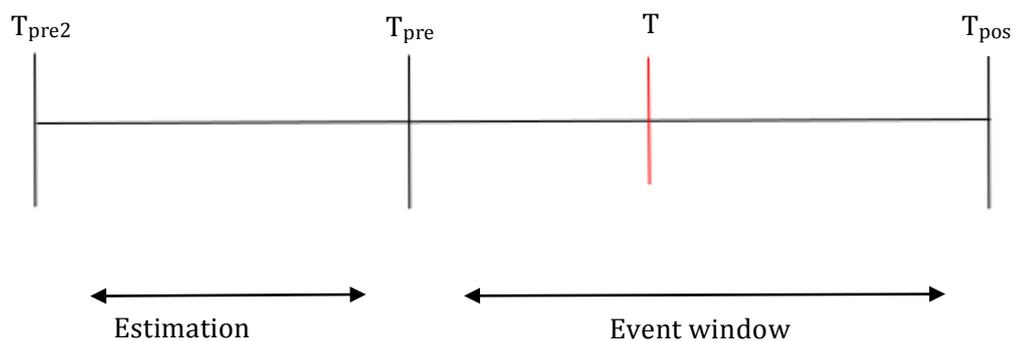

Figure 1. Sketch of time intervals on the M&A event time line.

In Fig.1,

$T_{pre2}$ = the first date of the estimation window

$T_{pre}$ = the first day used to calculate abnormal returns

$T$ = the event date

$T_{post}$ = the last date used in the calculation of abnormal return.

The estimation window chosen for the investigation takes 89 days prior to the event date as $T_{pre2}$ and 30 days prior to the event date as $T_{pre}$. Using the return on the stock and the corresponding return on the market previously calculated for each index in this window, a regression is created where alpha represents the intercept, and beta represents the slope.

From these parameters, an abnormal return can be defined as:

$$E_{jt} = R_{jt} - (\beta_{jt}R_{mt} + \alpha_{jt}) \tag{4}$$

Using these values generated for each stock *j*, the average abnormal return (*AAR*) is calculated for the pertinent event window. In this investigation, the event window takes 30 days prior to the first announcement and 10 days after. This provides the aggregate value of an abnormal return for all the stocks on day *t*.

The *AAR* is given by:

$$AAR_t = \frac{1}{N}\sum_{i=1}^{N} E_{jt} \tag{5}$$

where,

*N* = the number of stocks.

The t-test is then carried out on these *AAR* values generated for the event window in order to determine the significance of the abnormal return at time *t* and to observe which hypothesis has to be accepted within the t-statistics. The latter is computed with

$$t_{AARt} = \sqrt{N}\,\frac{AAR_t}{\sigma_{ARt}} \tag{6}$$

where,

$\sigma_{ARt}$ = the standard deviation of abnormal returns at time *t*, calculated from

$$\sigma_{ARt} = \sqrt{\frac{1}{N-1} \sum_{i=1}^{N} (E_{jt} - AAR_t)^2} \qquad (7)$$

The cumulative average abnormal return (*CAAR*) is another indicator of insider trading, also calculated for *N* days in the event window. This summates the previous daily average abnormal residuals. It can be used to examine the aggregate effects of abnormal returns. The *CAAR* is defined as:

$$CAAR = \frac{1}{N} \sum_{t=1}^{T} CAR_j \qquad (8)$$

where,

$CAR_j$ = the cumulative abnormal return for stock *j* is defined as:

$$CAR_j = \sum_{t=T_1+1}^{T_2} AR_{jt} \qquad (9)$$

## 5. Data and discussion

The empirical results from the estimation of abnormal returns of stock price data from the London Stock Exchange around merger announcements from two periods over the considered sample follow. Recall that the first period relates to the years 2008 to 2012; the second period relates to the years 2015 to 2019, thus separated by the Financial

Services Act 2012. Analysis of these returns relative to the announcement date and their statistical significance are provided here below as well as a comparison between the periods, with connexions to previous studies, and are related to asset pricing theories, recalled here above.

In Supplementary Files, Table 3 (left hand side, LHS) and Table 3 (right hand side, RHS) report the market model statistics for the aggregated sample for the event window, $t = -30$ to $t = 10$ for the period 2008-2012 and 2015-2019 respectively. The data is displayed in Figs. 2-4.

### 5.1. Time interval 2008-2012; before the Financial Services Act 2012

In the case of normal stock prices when there are no causes for unusual movements in prices, *AAR* and *CAAR* values are expected to fluctuate around zero. Consequently, in the event of a release of new information, a spike in both *AAR* and *CAAR* is anticipated. Thus, it can be presumed that information leakage will cause positive *AAR* values at $t = 0$ and accordingly lead to a build-up in *CAAR* (Keown and Pinkerton, 1981).

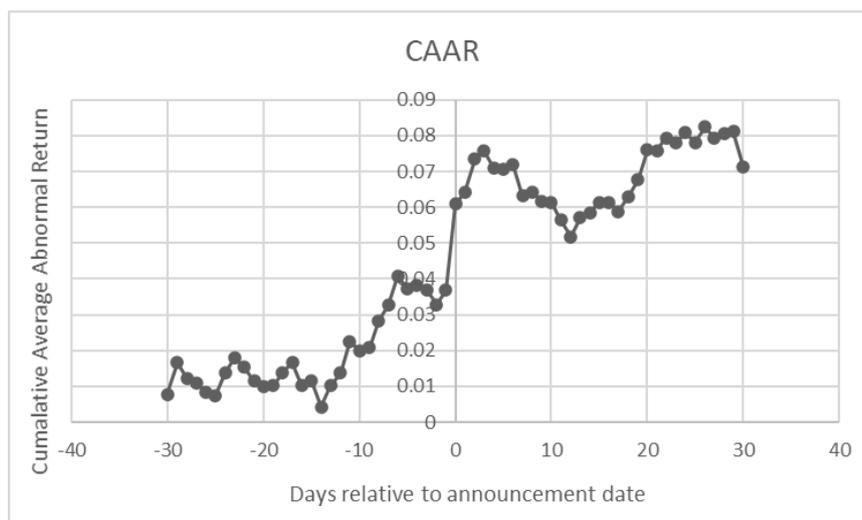

*Figure 2. Cumulative Average Abnormal Return (CAAR) for the period 2008 - 2012.*

Figure 2 shows the *CAAR* values of the run-up to merger announcements whilst under regulation by the FSA. A global upward trend is observed over the "first" 40 days; from day -30 before the M&A announcement date a positive *CAAR* value is observed. From an EMH and/or RW theory, *CAAR* is expected to fluctuate around zero if there are no unusual movements in the market. Thus, the lack of negative *CAAR* values observed in this period are likely explained by a leakage of information. This leakage can be assumed to have occurred before the 30 days prior to the announcement date in this event window.

Fig. 2 shows that before day -12, *CAAR* remains below 0.020. A dip in *CAAR* on -14 day is soon followed by an additional build-up which could be related to a leakage of information as the *CAAR* still fluctuates near zero prior to this. After day -12, *CAAR* begins to build up and by day -6, *CAAR* is recorded at 0.041 height, before the event date. This suggests an increased leakage of information in the market before the announcement. After day -6, the *CAAR* value starts to fall until the event date in which price (more fully) incorporates the information, as in EMH, and significantly increases. The slight fall in *CAAR* just prior to the event date confounds the explanation of insider trading and thus shows that insider trading may not actually be present here. The day just prior to the announcement, *CAAR* is observed at 0.037 as compared to 0.061 on the announcement date. Consequently, by this date 60.7% of the market reaction to the announcement has already occurred, thereby measuring the EMH level.

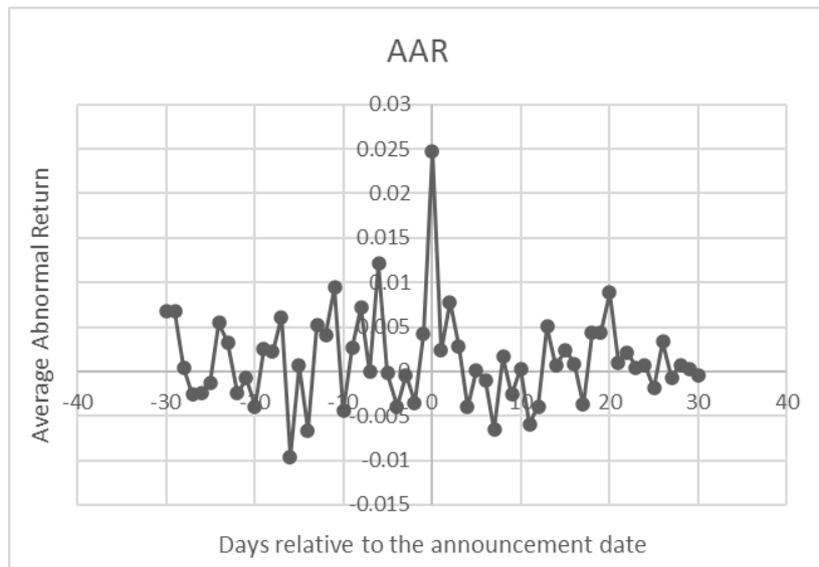

*Figure 3. Average Abnormal Return (AAR) for the period 2008 - 2012.*

Fig. 3 displays the *AAR* values in the event window. It can be observed in the run-up to the merger announcement, that *AAR* is positive for 20 days out of 30. A positive *AAR* suggests insider trading; however, the positive *AAR* values only make up 66% in the run-up to the event, such that it cannot be deemed a significant statistic. Further to this, negative *AAR* is observed on days -2 and -4 relative to the announcement date which may suggest a mere volatility in stock prices due to the RW theory rather than the explanation of insider trading. It is observed that day on -1 in relation to the announcement date abnormal returns begin to rise again. This increases by 0.008; it makes up for 27.5% of the increase in abnormal return from day -2 to day 0. Consequently, this would suggest a last-minute leak of information in the market.

Statistical testing carried out on the *AAR* values produced t-statistics for the days between -30 and +10 relative to the announcement date. These statistics determine whether the null or alternative hypothesis is accepted. The t-test carried out on the data from the period 2008 to 2012 shows that there are significant abnormal returns from as far back as 29 days prior to the announcement date.

Though significance is found, this is not consistent through the rest of the event window as abnormal returns were only significant before the announcement on days -29, -17, -14, and -11. Although the findings could suggest that insider trading occurs up to 29 days before the announcement, this would not be an appropriate conclusion to make. Indeed, it would also be expected that in the immediate run-up to the announcement, abnormal returns would also be significant. As this was not the case, insider trading on merger announcements for the period 2008 to 2012 is not considered to be significant. Subsequently, the null hypothesis is accepted for this period.

**5.2. Time interval 2015-2019; under regulation by the Financial Conduct Authority**

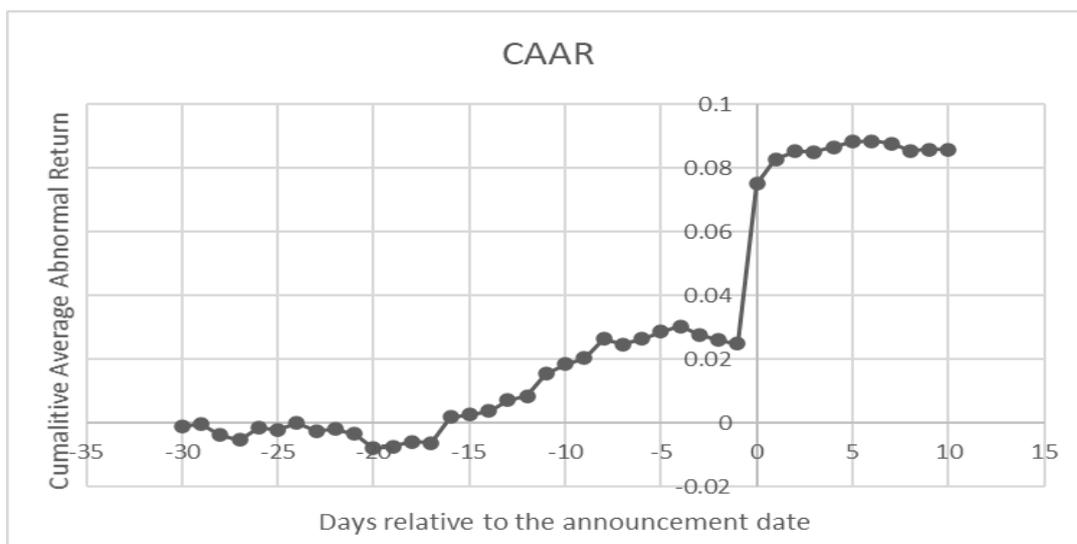

*Figure 4. Cumulative Average Abnormal Return (CAAR) for the period 2015 - 2019.*

Fig. 4 shows the trend of CAAR values from the period 2015-2019 thus after the implementation of the Financial Services Act 2012. The CAAR values show an increasing trend throughout the event window; this is not consistent through the lead up to the announcement. It can be observed that after first becoming positive on day -24, CAAR remains positive after -19 days relative to the announcement date. Further to this, on day -16, it can be noticed that a build-up in CAAR starts to emerge: it increases from 0.011 to 0.077 from day -16 to day 0. This markedly suggests a leakage of information where individuals have traded illegally on this.

Dissimilar to most studies which have found insider trading to be significant, our present findings show that the earlier build-up in CAAR that began on day -16 starts to fall on day -4 relative to the announcement date. In previous studies, CAAR is usually seen to continually increase until the announcement date if insider trading is present. Therefore, the present findings provide evidence that insider trading is in fact not present in the market in this instance. On day -4, the CAAR value is 41.5% of that on the announcement date; thus, it can be suggested that although it may not be significant, insider trading and leakage of information still occur prior to the merger announcements. The build-up that emerged on day -16 relative to the announcement date can be a sign of slight information leakages. On the announcement date, the security reaches the correct price in which this new information is incorporated. On the day before the announcement date, the CAAR value is 0.025 compared to 0.077 on the event date. Consequently, on the day prior to the announcement, only 32.5% of the market reaction had already occurred.

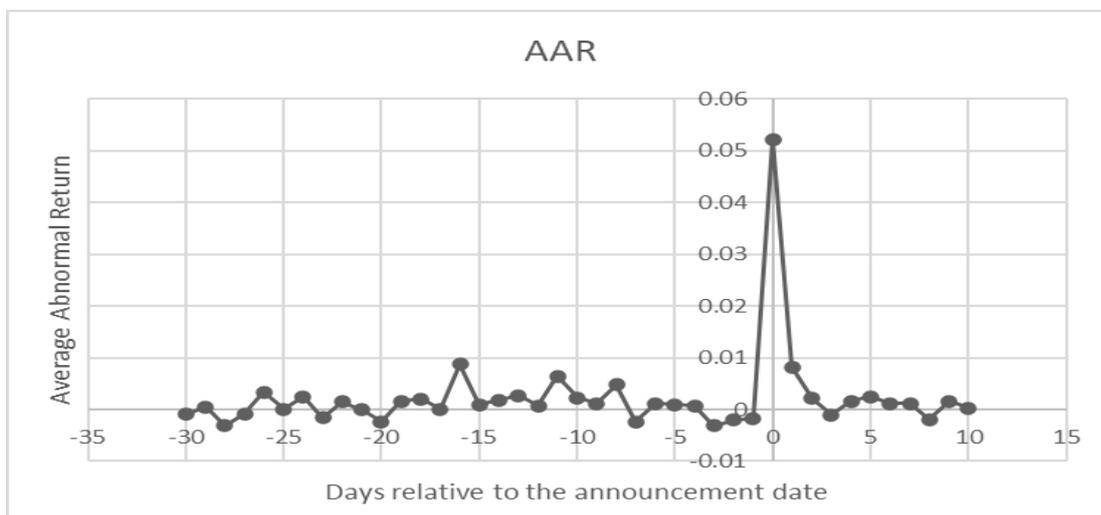

*Figure 5. Average Abnormal Return (AAR) for the period 2015 - 2019.*

Alike the previous period, out of the 30 days prior to the announcement date in the event window, only 10 of these have a negative AAR value. This means 66% of the abnormal returns are positive; this suggests a slight leakage of information. Despite this, the analysis of the *AAR* values in the days leading up to the announcement shows no sign of information leakage or anticipation of the announcement in the market. From day -6 onwards, *AAR* actually starts to decrease and is negative for the days -3 to -1 relative to the announcement date. In spite of the spikes in *AAR* values on days -16, -11 and -8, this would not indicate to outsiders that news have to be expected. Thus, the pattern is not enough to show any leakage of information prior to the announcement; the combination of these inconsistencies suggests little if no insider trading in this period. Subsequently, from Fig. 5, it can be assumed that there is no leakage of information as the price reaches the correct price on the announcement date with little contribution to this prior to the event date.

After conducting a t-test on the daily average abnormal return, there is no evidence to suggest that insider trading is present before an M&A announcement in this period. Abnormal returns were only found to be significant for days 0 and -8. This might display insider trading 8 days before the M&A information is released, but this could also be a consequence of other signals, for example any disposals carried out previous to the merger. Alongside this, significance was not consistent in the run-up to the announcement. Furthermore, the graph presenting *AAR* shows that there is only a sizeable increase in stock price on the event date. From the t-test of this period, the null hypothesis is also accepted.

**5.3 Discussion**

Following the examination of both periods, it is determined that the null hypothesis has to be accepted suggesting no statistical significance of abnormal returns. Thus insider trading has not been found to be present for both periods. However, differences regarding the presence of abnormal returns between the two periods still prevail.

In the analysis of both periods, it is noticed that a fall in *AAR* and *CAAR* values occurs just prior to the announcement date. This suggests that insider trading is insignificant in both periods and does not occur in the run-up to merger announcements in the UK. The insignificant returns can be explained as by Heinkel and Kraus (1987) who determined a statistical insignificance to be resultant to price fluctuations, finding that insiders do not outperform outsiders. Nonetheless, in terms of how effective the Financial Services Act 2012 has been in reducing insider trading, the present results show signs of reduced abnormal returns, whence reduced leakage of information prior to merger announcements. In comparison to the 2008-2012 period, the 2015-2019 period illustrates lower *AAR* values in the immediate run-up to merger announcements; this is measured at a height on day -8 relative to the announcement date at about 0.005 as compared to 0.012 on day -6 relative to the announcement date in the first period.

Thus, the results show that the Financial Services Act 2012 has been successful in reducing insider trading in the UK as the extent of abnormal returns before a merger announcement has lessened. It is also observed that the *AAR* values in the former period are a lot noisier than that of the latter period. This can be explained by the findings of Kim et al. (2019) who found that the absence of regulation to cause the exploitation of information, - and leads to noisier stock prices. Therefore, the tightening of regulation from the Financial Services Act 2012 can be determined to be effective in reducing noise in stock prices. Further improvement in regulation is observed in the 30 days before the announcement date where abnormal returns are only significant for 1 day in the 2015-2019 period compared to 4 days of significant abnormal returns in the former period.

This shows that the FCA is more efficient than the FSA in controlling the leakage of information and preventing individuals from abusing asymmetry. Despite this improvement, a build-up in *CAAR* from 16 days before the event date can be observed for the 2015-2019 period, suggesting that information leakage still occurs before the announcement. Subsequently indicating further room for improvement in additional control of insider trading by the FCA.

In terms of legislation, differences in abnormal returns between the two examined periods can be determined to be solely attributable to the Financial Services Act 2012. Further to this, since the Financial Services Act 2012, the Bank of England and Financial Services Act (2016) was created as another amendment. Regarding the FCA, the Bank of England and Financial Services Act (2016*)* only mentions the appointment of a chief executive to the regulatory body, with other adjustments to illegal money lending, money laundering, transformer vehicles, and pensions. With little focus on the FCA and insider dealing, it can be easily assumed that the Financial Services Act 2016 would have little effect on the extent of insider trading in the UK.

Despite the observation that in both periods *CAAR* begins to build up around 15 days prior to the announcement and reaches around half the *CAAR* value on the announcement date by day -1 relative to the event, some presence of insider trading and abnormal returns are not found to be significant. Thus, we differ with the majority of previous studies where significant abnormal returns are observed from around 7 days prior to the announcement date; in the present findings no significant abnormal returns were found in the immediate run-up to the merger announcements. The differences in results from the present study to previous ones, mostly for US markets, can be explained by the differences in executive characteristics between the UK and the US which indicate that UK executives have a lower need to diversify whence are more likely to hold acquired shares. The study by Keown and Pinkerton (1981), mentioned in Section 3, shows significant insider trading, - in the US, that conflicts with the present findings; this can also be explained by the executive differences but also by the differences in insider dealing regulation between the two countries. A thorough discussion of such differences is outside the frame of the present report. Nevertheless, one can show some satisfaction in such differences, since these characteristics suggest that executives in the UK have less motivation to trade illegally on private information than those in the US. In this respect, one explains the less imposing regulators found in the UK compared to those in the US. The more lenient regulators in the UK could be a result of the concept that insider trading is necessary to reduce the amount of asymmetrical information between insiders and outsiders, so further tightening of regulations could bring unintended negative consequences to shareholders (Chen et al., 2018).

Assessing the present results with the EMH, an association can be made with a study conducted by Del Brio et al. (2002) where insider trading in the Spanish stock market was examined; it was found that insiders would earn abnormal profits and outsiders would fail to obtain these profits when mimicking insiders, supporting the semi-strong form efficiency. In both periods examined in our study, the stock prices have adjusted to the information by the announcement date. So, alike the study by Del Brio et al. (2002), the semi-strong form EMH is supported by the present findings as past prices and public information are incorporated in the stock prices by the announcement date. Prior to the announcement date, the private information is not represented in stock price. Subsequently, the strong form EMH is invalid in this case; this somewhat surprisingly suggests that the insider trades so present in the run-up to M&A announcements are not informative.

## 6. Conclusion

Let us conclude having examined the effectiveness of the Financial Services Act 2012 in regulating insider trading. We utilise the event study methodology to assess abnormal returns in the run-up to the first announcement of mergers. Two samples of abnormal returns have been examined over two periods, either under regulation by the FSA or by the FCA. The samples are based on stock price data from the London Stock Exchange over 2008-2012 and 2015-2019, respectively. The results indicate that abnormal returns are reduced after the implementation of the Financial Services Act 2012; prices are also found to be noisier in the preceding period. Abnormal returns are found in the run-up to the first announcement of mergers in the 2015-2019 period, even though they are not considered to be statistically significant.

Concluding from such findings, in this study, it can be determined that through the enforcement of the Financial Services Act 2012 and the subsequent succession of the FSA by the FCA, insider trading is reduced. In spite of an absence of significant abnormal returns under the regulation of both the FSA and the FCA, it is still established that under the regulation of the FCA, abnormal returns are also reduced. Thus, despite the ongoing pressures from critics of the FCA stating their poor regulation of insider trading, our study concludes that there has been an improvement since the FSA, so the FCA has been effective thus far. The lack of abnormal returns in the 2015-2019 period suggests that no further tightening of regulation is practically required in the UK.

**6.1 Limitations**

We consider that there are 3 limitations in this study. The first limitation involves the sample used in this study: the numbers of merger announcements used in the two periods are unbalanced. The sample of merger announcements used in the 2008-2012 period consists of just under half of those used for the 2015-2019 period. The uneven group sizes may lead to a disproportionate extrapolation of data that generates misrepresentative results. The smaller group size from the first period was the result of a difficulty in gathering data from this time interval, when most of the accessible data only involves merger information from more recent years. To improve on this in the future, systems with large merger and acquisition databases could be used such as Thomson Reuters EIKON. This database was unavailable during the present study.

The second limitation regards the event window size and timing. Upon the analysis of the results, specifically from the 2008-2012 period, it is recognised that in the whole window, *CAAR* is positive. This would suggest that information is sometimes leaked before 30 days prior to the announcement.

Finally, the last limitation involves the assumed relationship between tightening regulations, and insider trading or abnormal returns. Indeed, we have assumed that these are equivalent words. Maybe they are not.

### 6.2 Further research

From the final limitation, suggestions for further research would involve a more in-depth examination of the relationship between regulation and insider trading (Rees, 1995). This would be done through the examination of trading volumes; it would be expected that insider trading would increase this.

---

We confirm that this article contains a Data Availability Statement (see Section 4 Methodology) : The data is based on information on M&A involving UK companies, gathered from the UK Office for National Statistics.

We confirm that we include how available data can be (and was) obtained

---

Acknowledgements

This paper is based on a dissertation by RP submitted for a degree fulfilment at ULSB. RP thanks the whole faculty for their teaching and concern.

# Bibliography

Agarwal, M. and Singh, H. (2006) Merger announcements and insider trading activity in India: an empirical investigation. *Investment Management and Financial Innovations*, *3*, pp.140-154.

Anker, G. (2013) 'We will be different to the FSA says new regulator, the FCA, as regime changes today', *MoneySavingExpert,* 1 April. Available at:

https://www.moneysavingexpert.com/news/2013/04/we-will-be-different-to-the-fsa-says-regulator-as-regime-changes-today/

Aspris, A., Foley, S., and Frino, A. (2014) Does insider trading explain price run-up ahead of takeover announcements?. *Accounting & Finance*, *54*(1), pp. 25-45.

Baesel, J.B. and Stein, G.R. (1979) The value of information: Inferences from the profitability of insider trading. *Journal of Financial and Quantitative Analysis*, *14*(3), pp. 553-571.

Bainbridge, S.M. (2019) The Law and Economics of Insider Trading 2.0. *Forthcoming in Encyclopedia of Law and Economics (2$^{nd}$ edition 2020)*, pp.19-01.

Bank of England and Financial Services Act 2016, c. 14. Available at: http://www.legislation.gov.uk/ukpga/2016/14/contents/enacted

BBC News (2018) 'Banking reform: What has changed since the crisis?', 4 February. Available at: https://www.bbc.co.uk/news/business-20811289

Bhattacharya, U. and Daouk, H. (2002) The world price of insider trading. *The Journal of Finance*, *57*(1), pp.75-108.

Bhattacharya, U. and Spiegel, M. (1991) Insiders, outsiders, and market breakdowns. *The Review of Financial Studies*, *4*(2), pp.255-282.

Binder, J. (1998) The event study methodology since 1969. *Review of Quantitative Finance and Accounting*, *11*(2), pp.111-137.

Binham, C. (2019) Suspect shares trades preceded one in 10 UK takeovers last year. *Financial Times*, 9 July. Available at: https://www.ft.com/content/493d0626-a242-11e9-a282-2df48f366f7d

Block, W.E. and Smith, T. (2016) The economics of insider trading: A free market perspective. *Journal of Business Ethics*, *139*(1), pp.47-53.

Bradley, M. and Seyhun, H.N. (1997) Corporate bankruptcy and insider trading. *The Journal of Business*, *70*(2), pp.189-216.

Bushman, R.M., Piotroski, J.D. and Smith, A.J. (2005) Insider trading restrictions and analysts' incentives to follow firms. *The Journal of Finance*, *60*(1), pp.35-66.

Carlton, D.W. and Fischel, D.R. (1983) The regulation of insider trading. *Stanford Law Review*, *35*(5), pp.857-895.

Chapman, B. (2018) FCA: City watchdog secures just 12 insider trading convictions in five years, *Independent*, 19 January. Available at:

https://www.independent.co.uk/news/business/news/fca-city-london-insider-trading-convictions-five-years-financial-conduct-authority-a8167486.html

Chen, Z., Guan, Y. and Ke, B. (2018) The Economic Consequences of Tightening Insider Trading Regulation: Evidence from Hong Kong. *Available at SSRN 3237234*.

Criminal Justice Act 1993, c. 36. Available at:

http://www.legislation.gov.uk/ukpga/1993/36/part/V

Darrat, A. F., and Zhong, M. (2000) On testing the random-walk hypothesis: a model-comparison approach. *Financial Review*, *35*(3), pp. 105-124.

Del Brio, E.B., Miguel, A. and Perote, J. (2002) An investigation of insider trading profits in the Spanish stock market. *The Quarterly Review of Economics and Finance*, *42*(1), pp. 73-94.

Doffou, A. (2003) Insider trading: a review of theory and empirical work. *Journal of Accounting and Finance Research*, *11*(1). Available at SSRN:

https://ssrn.com/abstract=1028898

Durnev, A., Morck, R., Yeung, B. and Zarowin, P. (2003) Does greater firm-specific return variation mean more or less informed stock pricing?. *Journal of Accounting Research*, *41*(5), pp. 797-836.

Dye, R.A., (1984) Inside trading and incentives. *Journal of Business*, *57*(3), pp. 295-313.

Fama, E.F. (1970) Efficient capital markets: A review of theory and empirical work. *The journal of Finance*, *25*(2), pp.383-417.

Fama, E.F., Fisher, L., Jensen, M.C. and Roll, R. (1969) The adjustment of stock prices to new information. *International Economic Review*, *10*(1), pp.1-21.

Fernandes, N. and Ferreira, M.A. (2009) Insider trading laws and stock price informativeness. *The Review of Financial Studies*, *22*(5), pp.1845-1887.

Financial Conduct Authority (2016) Insider dealers sentenced in Operation Tabernula trial. Available at:

https://www.fca.org.uk/news/press-releases/insider-dealers-sentenced-operation-tabernula-trial

Financial Services Act 2012, c. 21. Available at:

http://www.legislation.gov.uk/ukpga/2012/21/contents/enacted

Financial Services and Markets Act 2000, C. 8. Available at:

http://www.legislation.gov.uk/ukpga/2000/8/contents

Finnerty, J.E. (1976) Insiders and market efficiency. *The Journal of Finance*, *31*(4), pp. 1141-1148.

Fishman, M.J. and Hagerty, K.M. (1992) Insider trading and the efficiency of stock prices. *The RAND Journal of Economics*, *23*(1), pp. 106-122.

Fox, M.B. (1999) Retaining mandatory securities disclosure: Why issuer choice is not investor empowerment. *Virginia Law Review*, *85*(7), pp. 1335-1419.

Hall, O. (2013) 'Why the FSA was split into two bodies', *Financial Times Advisor*, 8 May. Available at: https://www.ftadviser.com/2013/05/08/regulation/regulators/why-the-fsa-was-split-into-two-bodies-SX5toVpnEQtBbYNlcUC9xJ/article.html

Heinkel, R. and Kraus, A., (1987). The effect of insider trading on average rates of return. *Canadian Journal of Economics*, *20*(3), pp. 588-611.

Hu, J. and Noe, T.H. (1997) *Insider trading, costly monitoring, and managerial incentives* (No. 97-2). Working Paper.

Jaffe, J.F. (1974) Special information and insider trading. *The Journal of Business*, *47*(3), pp. 410-428.
34

Jain, P. and Sunderman, M.A. (2014) Stock price movement around the merger announcements: insider trading or market anticipation?. *Managerial Finance, 40*(8), pp. 821-843.

Jensen, M.C. (1978) Some anomalous evidence regarding market efficiency. *Journal of Financial Economics*, *6*(2/3), pp. 95-101.

Jovanovic, F. (2018) A comparison between qualitative and quantitative histories: the example of the efficient market hypothesis. *Journal of Economic Methodology*, *25*(4), pp. 291-310.

Kendall, M.G. and Hill, A.B. (1953) The analysis of economic time-series-part i: Prices. *Journal of the Royal Statistical Society. Series A (General)*, *116*(1), pp. 11-34.

Keown, A.J. and Pinkerton, J.M. (1981) Merger announcements and insider trading activity: An empirical investigation. *The Journal of Finance*, *36*(4), pp. 855-869.

Kim, D., Ng, L., Wang, Q., & Wang, X. (2019) Insider Trading, Informativeness, and Price Efficiency Around the World. *Asia‐Pacific Journal of Financial Studies*, *48*(6), pp. 727-776.

Klock, M. (1994) The stock market reaction to a change in certifying accountant. *Journal of Accounting, Auditing & Finance*, *9*(2), pp. 339-347.

Kyriacou, K., Luintel, K.B. and Mase, B. (2010) Private information in executive stock option trades: evidence of insider trading in the UK. *Economica*, *77*(308), pp. 751-774.

Laffont, J.J. and Maskin, E.S. (1990) The efficient market hypothesis and insider trading on the stock market. *Journal of Political Economy*, *98*(1), pp. 70-93.Lakonishok, J. and

Lee, I. (2001) Are insider trades informative?. *The Review of Financial Studies*, *14*(1), pp. 79-111.

Lin, J.C. and Howe, J.S. (1990) Insider trading in the OTC market. *The Journal of Finance*, *45*(4), pp. 1273-1284.

Lorie, J.H. and Niederhoffer, V. (1968) Predictive and statistical properties of insider trading. *The Journal of Law and Economics*, 11(1), pp. 35-53.

Ma, Y., Sun, H.L. and Yur-Austin, J. (2000) Insider trading around stock split announcements. *Journal of Applied Business Research, 16*(3). pp. 13-25.

Manne, H.G. (1966) *Insider Trading and the Stock market*. New York: The Free Press.

Manove, M. (1989) The harm from insider trading and informed speculation. *The Quarterly Journal of Economics*, *104*(4), pp. 823-845.

Practical Law (2018) Banking regulation in the UK: overview. Available at: https://uk.practicallaw.thomsonreuters.com/w-008-0211?transitionType=Default&contextData=(sc.Default)&firstPage=true&bhcp=1

Rees, B. (1995) Financial analysis (No. 2nd). *Hertfordshire: Prentice Hall*.

Richardson, S., Teoh, S.H. and Wysocki, P.D. (2004) The walk-down to beatable analyst forecasts: The role of equity issuance and insider trading incentives. *Contemporary Accounting Research*, *21*(4), pp. 885-924.

Rosen, R.J. (2006) Merger momentum and investor sentiment: The stock market reaction to merger announcements. *The Journal of Business*, *79*(2), pp. 987-1017.

Rozeff, M.S. and Zaman, M.A. (1988) Market efficiency and insider trading: New evidence. *Journal of Business*, *61*(1), pp. 25-44.

SEC (2004) 'SEC charges Jeffrey K. Skilling, Enron's former president, Chief Executive Officer and Chief Operating Officer, With Fraud'. Available at:

 https://www.sec.gov/news/press/2004-18.htm

SEC (2011) 'SEC Obtains Record $92.8 Million Penalty Against Raj Rajaratnam.' Available at:  https://www.sec.gov/news/press/2011/2011-233.htm

SEC (2013) '2013 Insider trading policy.' Available at:

 https://www.sec.gov/Archives/edgar/data/25743/000138713113000737/ex14_02.htm

Seyhun, H.N. (2000) *Investment intelligence from insider trading*. MIT Press, Cambridge, MA.

Wijesinghe, P. (2018) Criminal and Civil Offences of Insider Dealing: UK Perspective. *Available at SSRN 3167197*  http://dx.doi.org/10.2139/ssrn.3167197

Wuttidma, C.P. (2015) *Aggregate insider trading activity in the UK stock and option markets* (Doctoral dissertation, Brunel University, London).